# Prediction of mechanical properties of non-equiatomic high-entropy alloy by atomistic simulation and machine learning

Liang Zhang[1,2,†,*], Kun Qian[3,†], Björn W. Schuller[4,5], Cheng Lu[6], Yasushi Shibuta[2,*], Xiaoxu Huang[1]

[1] College of Materials Science and Engineering, Chongqing University, Chongqing, 400044, China

[2] Department of Materials Engineering, The University of Tokyo, Tokyo, 113-8656, Japan

[3] Educational Physiology Laboratory, The University of Tokyo, Tokyo, 113-0033, Japan

[4] Department of Computing, Imperial College London, London, SW72AZ, UK

[5] Department of Computer Science, University of Augsburg, Augsburg, 86159, Germany

[6] Faculty of Engineering and Information Science, University of Wollongong, Wollongong, NSW 2522, Australia

[†] These authors contributed equally to this work.

[*] Corresponding author. Email: liangz@cqu.edu.cn (L.Z.); shibuta@material.t.u-tokyo.ac.jp (Y.S.)

## Abstract

High-entropy alloys (HEAs) with multiple constituent elements have been extensively studied in the past 20 years due to their promising engineering application. Previous experimental and computational studies of HEAs focused mainly on equiatomic or near equiatomic HEAs. However, there is probably far more treasure in those non-equiatomic HEAs with carefully designed composition. In this study, molecular dynamics (MD) simulation combined with machine learning (ML) methods were used to predict the mechanical properties of non-equiatomic CuFeNiCrCo HEAs. A database was established based on a tensile test of 900 HEA single-crystal samples by MD simulation. We investigated and compared eight ML models for the learning tasks, ranging from shallow models to deep models. It was found that the kernel-based extreme learning machine (KELM) model outperformed others for the prediction of yield stress and Young's modulus. The accuracy of the KELM model was further verified by the large-sized polycrystal HEA samples.

# Introduction

Excellent mechanical properties are a prerequisite for the development of advanced structural materials. Combining high strength and good ductility is highly desired, yet challenging for conventional structural materials. In the past few decades, several materials with concurrent high strength and good ductility have been developed[1-3], as exemplified by super-alloys, dual-phase steels, nanotwinned materials, gradient materials, and bulk-metallic glasses. These materials are based on one main element, while additional elements are added to improve specific properties. For example, the mechanical strength of a material can be increased by alloying followed the classical hardening mechanisms such as solid solution, precipitation, dispersion, and second phase strengthening. A distinctly different concept of "high-entropy alloys" (HEAs), in which metals are mixed by five or more elements in equiatomic or near-equiatomic proportions, was proposed by Yeh and co-workers[4] and independently by Cantor and co-workers[5] in 2004. After that, HEAs have been extensively studied by researchers due to their promising properties[6,7]. Compared to conventional alloys which contain one and rarely two base elements, HEAs have proven to show many superior mechanical properties. These advanced properties include excellent strength, exceptional ductility and fracture toughness at cryogenic temperatures, superior mechanical performance at high temperatures, super-paramagnetism, and superconductivity[8]. Some recent high-profile studies have shown that the well-designed HEAs exhibit superior mechanical properties that can overcome the longstanding conundrum of the strength–ductility trade-off in materials[9-11].

HEA is a large field with a countless number of new alloy systems. If five elements were randomly selected from a pool of 118 elements to form an alloy, the number of combinations could reach about 200 million using the combination theory. If only 38 transition metals were considered, the number of new alloys would still be as high as about 500,000[8]. Due to their considerable structural and functional potential as well as the richness of design, different combinations of the elements and their compositions need to be explored to understand the potentials of HEAs further. However, our understanding of these new alloys is still very limited and is still at an early stage. The equiatomic composition is usually the easiest access to a new alloy system, and the HEAs with equiatomic compositions were intensively investigated by many of the previous research works. However, the equiatomic HEAs probably do not possess the best combination of properties. For example, Yao *et al.*[12] reported a novel, single-phase, non-equiatomic $Fe_{40}Mn_{27}Ni_{26}Co_5Cr_2$ (at.%) HEA with exceptional phase stability and excellent tensile ductility. Fu *et al.*[13] found that the bulk nanocrystalline $Co_{25}Ni_{25}Fe_{25}Al_{7.5}Cu_{17.5}$ (at.%) HEA can exhibit a compressive yield strength of about 1.8 GPa, which is dramatically higher than the yield strength of most previously reported fcc structured HEAs. As suggested by Yeh, the pioneer of HEAs, 'there is probably far more treasure in those non-equimolar alloys with carefully designed composition and tailored microstructures'[14]. A good method is to start with equiatomic HEA and then change the composition of the consisted elements according to the target properties for the design of a new non-equiatomic HEA system. However, an empirical trial-and-error approach requires a large amount of time and cost of designing, it is impossible to evaluate such huge numbers of the HEA system in the available time by traditional methods, especially for experimental methods. High-throughput computational materials design integrated with advanced machine learning (ML) methods provide new tools for exploring the vast space of possible materials and finding the required properties. This field represents an emerging and dynamic scientific activity that has contributed significantly to the discovery of new materials[15-19]. The basic strategy is to combine the high-throughput computation results or materials databases with intelligent data analysis techniques, giving full play to its advantages to seeking target performance. The composition design of HEA provides an ideal platform for developing, validating, and applying these new techniques.

In this work, molecular dynamics (MD) simulation was combined with advanced ML methods to predict the mechanical properties of HEAs based on their element information; the schematic diagram is shown in Fig. 1. Firstly, the mechanical responses of 900 single-crystal CuFeNiCrCo samples with different compositions were obtained by MD simulations. The single-crystal samples cover specific crystallographic orientations by considering the anisotropy of the mechanical property of the material. Based on the MD results, we established a database describing the relationship between element composition and mechanical properties of the tested HEA samples. Secondly, the overall dataset was divided into train, development (dev), and test sets. We investigated and compared eight different ML models, including a deep neural network (DNN), extreme learning machine (ELM) and Kernel-based ELM (KELM), and support vector machine (SVM) among others. To determine the most suitable algorithm, the output features of ML, namely the yield stress and Young's modulus, are classified into 'Good' and 'Weak' based on a benchmark value. The KELM model outperformed others when dealing with a small database in this study. Finally, we constructed ten large polycrystal CuFeNiCrCo samples with different compositions. The well-trained ML model, based on single-crystal samples, was used to predict the mechanical responses of the polycrystal samples. The prediction from the KELM model was in good agreement with the calculated results from the MD simulation. We show in this contribution that computational study combined with ML methods is an efficient way to predict the mechanical properties of HEAs.

## Results

**Mechanical response of HEA single-crystal samples.** We first carried out MD simulations to test the tensile responses of the CuFeNiCrCo single-crystal samples. The interatomic potential[20] used in this study can accurately reflect the basic structural and physical characteristics of the five components, including the lattice constant, cohesive energy, and elastic modulus. Some important parameters are listed in Table S1 in the supplementary materials. In particular, the interatomic potential can accurately present a large variation in stacking fault energies of the constituent elements, both stable and unstable (Section S1). This is significant for the study of dislocation behavior and, therefore, the mechanical properties of the alloy system. The simulated samples with a face-centered cubic (fcc) phase has been confirmed as a stable structure in the previous experimental studies of the similar alloy system[8]. Here, we further examined the phase stability of the HEA samples by gradually raising the system temperature from 300 K to 3000 K (Section S2). It was found that the fcc structure was retained until the sample were heated up to a high temperature due to the thermal diffusion mechanism.

Fig. 2a shows the mechanical responses of the equiatomic HEA samples with different orientations under uniaxial tension at 300 K. Young's modulus for each crystal orientation is obtained from the linear portion of its corresponding stress-strain between 0 and 0.5% strain. Theoretically, the yield stress and the incipience of plasticity correspond to the nucleation of the first set of dislocations within grains[21]. Here, it is found that the dislocation nucleation event occurs near the peak of the stress-strain curve, as shown in Fig. 2b. Therefore, the maximum tensile stress is defined as the yield stress in this work. The mechanical response shows an elastic anisotropy of the HEA samples with different crystal orientations. The tension along [111] orientation shows the maximum Young's modulus of 268.96 GPa, while the [210] orientation has the minimum value of 169.07 GPa. Also, there are remarkable differences in yield stress for different orientations, the maximum yield stress is 15.14 GPa for [111] orientation, and the minimum value is 7.66 GPa for [110] orientation. Therefore, considering the anisotropy of the mechanical properties of materials can contain a wider range of information than simply considering the element composition when establishing a database, which is conducive to improving the reliability and applicability of an ML model in dealing with complex structural materials. The stress-strain responses of the non-equiatomic HEA samples with different orientations are presented in Fig. S3. For each orientation, the figure contains 100 results based on the random combination of the constituent elements. The results of the equiatomic HEA sample are also plotted for comparison. It is found that the yield stress and Young's modulus of the equiatomic HEA sample are at a medium level for all tested orientations. For example, Fig. 2c plots the stress-strain responses of five HEA samples with different element compositions along [110] orientation. The maximum Young's modulus is 199.5 GPa for $Cu_{21}Fe_{18}Ni_{27}Cr_{23}Co_{11}$ sample, and the minimum value is 183.5 GPa for $Cu_{31}Fe_{8}Ni_{10}Cr_{32}Co_{19}$ sample. The maximum yield stress of 9.78 GPa is observed for $Cu_{20}Fe_{33}Ni_{32}Cr_{10}Co_{5}$ sample, and $Cu_{26}Fe_{9}Ni_{9}Cr_{22}Co_{34}$ sample has the minimum value of 6 GPa. For the equiatomic HEA sample, the yield stress (7.66 GPa) and Young's modulus (192.3 GPa) are at a medium level among the five samples.

Two conclusions can be drawn from the above results. On the one hand, changing the proportions of elements can improve or decrease the mechanical properties of HEA samples, which provides an opportunity for achieving the required performance of HEA by optimizing its element composition. On the other hand, the high Young's modulus does not guarantee high yield stress. For the equiatomic HEA sample, the yield stress of [100] orientation is 18.26 GPa, which is more than twice as much as that of [110] orientation (7.66 GPa). However, Young's modulus of [100] orientation (118.7 GPa) is much lower than that of [110] orientation (192.27 GPa). For further illustration, Fig. 2d plots the yield stress ($E$) as a function of Young's modulus ($Y_s$) of different HEA samples under tension along [110], [210], and [821] orientations. The fitting lines guide the general trend of the point sets. The scattered points indicate that there is no direct correlation between $E$ and $Y_s$ of the HEA samples. The lower correspondence between Young's modulus and yield stress is mainly ascribed to the nonlinear elastic behavior of materials under strain, either 'elastic hardening' or 'elastic softening' during dynamic tension in the elastic stage[22]. The yield stress is usually small accompanied by the elastic softening, while a high yield stress value can be achieved when the elastic hardening occurs. The $E$-$Y_S$ relationships for other orientations are shown in Fig. S4. Although the yield stress generally increases with the increase of Young's modulus for most orientations, the correspondence between the two values is not prominent. The independence of Young's modulus and yield stress makes it necessary to predict the two mechanical properties by ML separately.

**ML models training and evaluation.** Based on the MD simulations on the tensile responses of 900 non-equiatomic single-crystal HEA samples, the quantitative relationship between element composition (input features) and the mechanical properties (*Ys* and *E*, output features) of HEA samples were obtained, and the database used for ML was constructed. In this section, we investigated eight ML models for the given tasks, ranging from shallow models to deep models. The underlying principle of the eight ML models are briefly introduced in Section S3.

The overall dataset was split into train, development (dev), and test sets, which occupies 60%, 20%, and 20% of the whole instances, respectively. In order to eliminate the effects of outliers, all the input features (the element composition) were standardized using a *z*-score transformation before being fed into the ML models. The feature vectors were scaled to a distribution having an arithmetic mean of zero and a variance of one. Let $\boldsymbol{x}(n)$ be the composition vector of the *n*-th element in the high-entropy alloy data, and we standardize the values of $\boldsymbol{x}(n)$ as: $\tilde{\boldsymbol{x}}(n)=[\boldsymbol{x}(n)-\mu_x]/\sigma_x$, where $\mu_x$ and $\sigma_x$ are the mean and the standard deviation of the vector $\boldsymbol{x}(n)$, respectively. We keep this information measured in the train set ($\mu_x$ and $\sigma_x$) and apply it to the dev and test sets. Considering the imbalanced characteristic of our proposed database (i.e., the number of instances belonging to each class is unequal among the database), we use the unweighted average recall (UAR) as our main metric to evaluate model performance (Section S4). The UAR metric is regarded as more suitable and rigorous than the weighted average recall (WAR, i.e., the accuracy) for evaluating a model's performance based on an imbalanced database[23]. On the other hand, the WAR, the sensitivity (Sens.), the specificity (Spec.), the precision (Prec.), and the F1-score are calculated as complementary evaluation metrics. The hyper-parameters of all ML models are tuned and optimized on the dev set based on the performance (UAR). We use a grid-search strategy to decide the optimal hyper-parameter for a specific ML model. The detailed information of the grid-search strategy and the procedure of hyper-parameter optimization are introduced in Section S5. When conducting the final evaluation, the data of the train and dev set are merged together to train the ML model within the optimized hyper-parameters, which is used to predict the output features (*Ys* and *E*) of the HEA samples in the test set.

Fig. 3a and b present the achieved results (UARs in [%]) of the eight ML models for the learning task of yield stress (Ys) and Young's modulus (E), respectively. The results of the complementary evaluation metrics are listed in Table S4. We can see that most of the results are above 85.0% of UAR, indicating that the ML models perform efficiently for both of the two tasks. For the task of yield stress, the best UAR achieved on the test set is reached by the NB model (86.4%), while for the task of Young's modulus, the KELM model takes the first place with a UAR of 92.2%. It is found that some simple ML models (e.g., NB, *k*-NN, LDA) can also show a good capacity in predicting the results; they perform well for both of the two tasks. It is reasonable to think that the relationship between the element composition (input features) and the mechanical properties (output features) of the HEA is sufficient to build ML models. Compared to the DNN model, a simple multi-layer perceptron structure without pre-training process in this study, the SAE model can learn more information inherited from the data itself in an unsupervised paradigm due to their pre-training process. Therefore, the SAE outperforms the general DNN for the two tasks on the test set. However, both of the two models have been constrained in this study, mainly due to the limited instance number of the small database for deep learning models. In contrast, ELM and its variant, KELM, are demonstrated to be more efficient in this study. Unlike the two deep models, the ELM and KELM models do not require to tune the parameters of the hidden nodes, which makes their training process much faster. Moreover, for a small size dataset in this study, ELM and KELM were able to execute efficiently while maintaining a fast training scheme. SVM is a popular and stable ML model that has been successfully applied to many tasks during the past decades. In this study, although SVM has shown some robustness and effectiveness, the results are not as good as ELM and KELM. It was indicated that, when using kernels, SVM is more likely to get sub-optimal solutions than KELM, which may lead to a negative result[24]. By comparing the UAR results as well as other complementary evaluation metrics of the above models, we think that the KELM model has a better overall performance for the two learning tasks, and it is selected as the preferred ML model in the subsequent prediction studies.

To further demonstrate the effectiveness of the KELM model, Fig. 3c and d plot the 180 predicted results on the test set by the KELM model. Firstly, the results of yield stress and Young's modulus from MD simulations are divided into the two categories: 'Good' and 'Weak', respectively. The red line is the benchmark line defined by the values of the equiatomic sample, and the value is set as one. The 180 results of the non-equiatomic samples are normalized according to the benchmark value. The points above the red line are classified into the 'Good' zone, while the points below the red line are classified into the 'Weak' zone. If the predicted result ('Good' or 'Weak') from the KELM model matches the MD result, the point is presented as blue, otherwise in red. It is found that most of the red points are located near the red line, implying that the KELM model gives a negative result only when the predicted value is too close to the benchmark value. This is more evident in the task of Young's modulus, as shown in Fig. 3d. Based on the above results, we may assume that if an upper and lower threshold was set for the benchmark value, the prediction accuracy of the ML model could be further improved, which again proves the efficiency of the KELM model regarding the learning tasks.

**Prediction of HEA polycrystal samples.** We construct large-size polycrystal HEA samples to verify further the reliability and applicability of the well-trained KELM model. The polycrystal sample is the combination of many single crystals with random orientations, which can better characterize the texture of real materials. As shown in Fig. 4a, the remarkable difference from the single-crystal sample is that the polycrystal sample contains complicated grain boundary networks. The sample is colored according to the CNA method, in which green atoms are in the grain with fcc structure, and blue atoms are in the grain boundary region with a disordered structure. For simplicity, the HEA tested polycrystal samples are marked as P0 to P10, where P0 is the sample with equiatomic composition, and P1 to P10 are ten non-equiatomic samples. Each polycrystal sample contains 16 grains with a mean grain size of 24.8 nm. The elements information of the samples is shown in Table 1. Fig. 4b plots the strain-stress response of the polycrystal samples with different compositions under uniaxial tension by MD simulation. The peak stresses of the samples are found between 4~5% strain depending on their element compositions. It is found that the yield stress and Young's modulus of the equiatomic samples are at a medium level comparing with the non-equiatomic counterparts. The result is consistent with the case of single-crystal samples, that is, changing the element composition can either improve or decrease the mechanical properties of the polycrystal HEA samples.

Based on the MD simulation, the results of yield stress and Young's modulus are classified into two groups. Meanwhile, the prediction results ('Good' or 'Weak') are given by the ML model. The comparison results of MD simulation and ML prediction are listed in Table 1. For yield stress, it was found that the ML model can give a correct prediction on nine tested samples with only P4 for an exception. The accuracy is consistent with the prediction of single-crystal samples (~90%). However, the ML model failed to predict Young's modulus of four (P2, P5, P9, and P10) out of the ten tested samples; the predictive accuracy is much lower than that in the case of the single crystal. This deviation is mainly attributed to the grain boundaries in the polycrystal sample. Experiments and simulations have adequately demonstrated that the presence of grain boundaries has a significant impact on the mechanical performance of materials[25-27], especially when the grain size drops to the nanocrystalline region (<100 nm)[28]. As aforementioned, the yield stress depends mainly on the nucleation of the first set of dislocations [21], and the stress required for dislocation nucleation at grain boundary is much lower than that required for nucleation of the single-crystal sample with defect-free structure[29,30], so the yield stress of polycrystal samples shows an overall decreasing trend when compared with the single-crystal samples. However, similar to the single-crystal sample, dislocation nucleation is still the main reason for polycrystal samples to yield under the current gain size. Take $Cu_{28}Fe_{13}Ni_{23}Cr_8Co_{28}$ as an example, Fig. 4c show the configuration of the sample at different deformation stages under tension. These snapshots are the cross-sectional view of the sample along [110] direction. We can see that, the yielding of the sample occurs at tensile strain between 4~5%, which corresponds to the initial dislocation nucleation from grain boundaries. After that, the system stress decreases rapidly with more dislocations nucleated from grain boundaries. However, it is shown that the configuration of the grain boundary network has not changed substantially near the yielding point, indicating that the presence of grain boundaries does not play an important role in the yield process of the polycrystal sample for current grain size, thus almost no impact on the yield stress task of ML.

On the other hand, Young's modulus is sensitive to the chemical composition and the intrinsic structure of materials. Simulations have shown that Young's modulus decreases with the increase of grain boundary volume fraction[31]. The single-crystal sample has a single fcc structure (green atoms), but for the polycrystal samples, the atoms at grain boundary regions contribute a new disordered structure (blue atoms). Considering the current computing ability of MD simulation, the mean grain size of the polycrystal sample is limited at tens of nanometers, in which the number of atoms at grain boundaries has a considerable proportion of all the atoms in the sample. Therefore, grain boundaries can have more prominent influence on the result of Young's modulus than that of yield stress. Since the ML model was trained and optimized based on single-crystal samples without considering the effect of grain boundary on Young's modulus, it is reasonable to see a deviation when predicting the polycrystal samples using the same model. However, the authors believe that if the mean grain size can be increased by orders of magnitude, the accuracy of the ML model will be further improved. Moreover, it is found that the red points (failed results by ML) are close to the benchmark line. If we give an upper and lower threshold of 2% on the reference value of the atomic sample, three of the four failed points (P5, P9, and P10) for Young's modulus can be predicted as correct, and the only failed point (P4) for yield stress can also be classified into correct. Therefore, a minor modification on the well-trained KELM model from the single crystal case can lead to a more favorable prediction result on the polycrystal sample, thus potentially significantly reducing the computing time.

## Discussion

MD simulation combined with ML methods was used to predict the mechanical properties of non-equiatomic CuFeNiCrCo HEA samples. Firstly, the mechanical properties of the HEA single-crystal samples under uniaxial tension along different crystallographic orientations were examined by MD simulations. It was found that the mechanical responses of the tested samples can change considerably depending on the element composition, indicating that there is a large space for improving the performance of HEA by optimizing the element composition. Secondly, the yield stress and Young's modulus were chosen as the learning targets, and we investigated and compared different ML models and algorithms, including shallow models (NB, LDA, *k*-NN, SVM, ELM, KELM) and deep models (DNN and SAE). By training, developing, and testing on 900 non-equiatomic HEA single-crystal samples, the most efficient ML model was identified and optimized. According to different evaluation metrics, the KELM model was found to give a prediction with high accuracy for both learning targets, and it was observed as the most appropriate model for the small database in this study. Finally, we constructed ten large HEA polycrystal samples with non-equiatomic composition and examined their mechanical properties by MD simulation. The well-trained KELM model was further used to give a prediction on the performance of polycrystal samples. The result shows that the prediction on yield stress is basically consistent with the simulation results, while the prediction on Young's modulus shows lower accuracy. The deviation of the prediction results is mainly ascribed to the presence of grain boundaries in the polycrystal sample. However, the benchmark value with a 2% threshold can greatly improve the prediction accuracy of the KELM model on polycrystal samples.

Like previous ML studies on mechanical or physical properties of materials[32-34], only limited output features were selected as learning targets in this work. It must be emphasized that, while yield stress and Young's modulus are important mechanical parameters affecting the strength of materials, many other important factors, such as work-hardening rate, ductility, toughness, *etc.* should be considered when developing new materials. One has to choose the appropriate targets according to the actual needs of the material design. The unexplored space of the element compositions for optimized properties of HEAs is still large, and multi-objective learning is an effective way to improve the comprehensive properties of materials. Nevertheless, we have developed a new method for the design and prediction of HEA properties by combining the atomistic simulation and ML approaches efficiently. That is, a certain number of material models and their parameters can be obtained by high-throughput computation, providing a database for ML. With the help of a well-trained ML model, we can evaluate the materials with target properties and then screen out the schemes which satisfy the given requirements. In particular, by considering specific material properties (*e.g.,* mechanical anisotropic in this study), it is possible to obtain high accuracy ML model based on a small database. This method will provide instructive guidance for the sample preparation at the experiment stage and will significantly accelerate the development of new HEA materials for engineering applications.

## Methods

**Computer simulation.** MD Simulations were performed using the parallel molecular dynamics package LAMMPS[35] with the embedded atom method (EAM) interatomic potentials developed for the CuFeNiCrCo HEA system[20]. The single-crystal HEA samples without initial structural defects were generated, and the length of each side of the cube sample is approximately 10 nm. The current model size is a benefit of the computational efficiency and ensures the observation of primary nucleation of dislocations in the grain. The large polycrystal samples were constructed using the Voronoi construction method containing 16 grains with random crystallographic orientations. The mean grain size is 24.8 nm, and the total number of atoms is 10,976,192. Atoms of the five elements are uniformly distributed in both single-crystal and polycrystal HEA samples. All simulation samples were constructed with a single fcc phase. The equilibrium structure of the HEA sample was obtained after an initial energy minimization and the followed system annealing procedure for 100 ps. The simulations were performed at the temperature of 300 K. Periodic boundary conditions were applied to all directions, atomic vibration and change in sample dimensions were allowed during the simulation. The fcc lattice structure of the samples was maintained after system annealing in an isobaric-isothermal (NPT) environment. In order to test the mechanical properties of HEA samples with different compositions, the uniaxial tension was applied in the Y direction while the pressure along the other two directions was kept at zero in the NPT ensemble. The deformation strain was set at a constant rate of $5 \times 10^8$ s$^{-1}$, and the timestep was set as 1 fs throughout the work. The system strain was derived from the positions of the periodic boundaries along the loading direction, and the system stress was attained by calculating the pressure of the entire system of atoms. All the atomic figures in this study are illustrated by the visualization tool Ovito[36], and the dislocation extraction algorithm (DXA)[37] is used to extract the dislocation lines in the sample during deformation.

**ML targets.** The mechanical properties of HEAs can be different from any of the constituent elements. The structure types and composition of the elements are the dominant factors for controlling the strength or hardness of HEAs[7]. Because of the wide composition range and the enormous number of alloy systems in HEAs, the mechanical properties of HEA can vary considerably. For metallic materials, the first dislocation nucleation event plays an essential role in determining the mechanical strength. Theoretically, it corresponds to the yield strength and represents the beginning of the plastic deformation of materials[21]. Also, the yield strength plays a significant role in the contribution of materials hardening mechanisms and hence can be used to guide the design and optimization of alloy composition. For example, Shen *et al.*[33] recently used an ML method to explore the ultrahigh-strength stainless steel by alloy design. In this study, yield stress ($Y_S$) is determined as the first target for ML. On the other hand, Young's modulus is an intrinsic mechanical property mainly determined by the elements of their constitutions and crystal structures. For the conventional one-element principal alloys, Young's modulus is mainly controlled by the dominant element. In contrast, for HEAs, Young's modulus can be very different from any of the constituent elements in the alloys[38]. For example, Wu *et al.*[34] used an ML method to predict Young's moduli of Ti alloys for the design of new products with bio-compatibility and low modulus. Here, we chose Young's modulus ($E$) as the second target for ML.

**Database.** The quantitative relationship between elements composition and the mechanical properties of HEA samples was obtained from MD simulations. In the ML task, the composition of the five elements CuFeNiCrCo was used as input feature to predict the mechanical response. The principle concept of HEAs is based on designing the alloys with multiple principal elements ranging from 5 to 35 atomic percent with a target to form single-phase solid solutions arising from high entropy of the system[4]. Therefore, when randomly assigning the proportion of the five elements, the upper limit of each element was controlled to be 35%, and the lower limit to be 5%. The mechanical properties, including the yield stress ($Y_S$) and Young's modulus ($E$) were set as output features, and the classification of the output values is the learning target. In particular, the traditional material design by ML is only based on the design of a pure mathematical algorithm, while the intrinsic structural and physical properties of the material are not considered. In this way, ML usually requires large data information and may result in uneven learning samples in terms of the special properties of materials. In this study, when establishing the database, we not only randomly changed the composition of the elements in the alloy system, but also considered an important physical index of materials, namely the anisotropy of the mechanical properties, when conducting the MD simulation. Here, we deliberately selected nine different crystallographic orientations to test the mechanical properties of HEA samples. These orientations are the representative orientations in the texture of materials, including the vertex position ([100], [110], [111]), edge position ([210], [211], [221]), and internal position ([321], [821], [876]) of the the inverse pole figure for material textures (Fig. 1d). MD simulations on each orientation contains a result of the equiatomic HEA sample and 100 results of the non-equiatomic HEA samples based on different elements compositions.

**ML algorithms.** There are three main paradigms in ML, i.e., supervised learning, unsupervised learning and reinforcement learning. Our task is to predict the mechanical properties (output features) of the HEA samples through element composition as the input features; hence, the supervised learning paradigm is selected. We investigate and compare eight ML models ranging from classic (shallow) models (i.e., Naïve Bayes (NB), linear discriminant analysis (LDA), *k*-Nearest Neighbour (*k*-NN), support vector machine (SVM), extreme learning machine (ELM), and kernel-based extreme learning machine (KELM)) to a deep neural network (DNN), and stacked auto-encoders (SAEs)). The underlying principle of the eight ML models are briefly introduced in Section S3.

# Data availability

The data that support the findings of this study are available in the supplementary materials and dataset. Additional data related to this paper may be requested from the corresponding authors upon reasonable request.

# References

1. T. Zhu, J. Li, Ultra-strength materials. *Prog.Mater. Sci.* **55**, 710 (2010).
2. E. Ma, T. Zhu, Towards strength–ductility synergy through the design of heterogeneous nanostructures in metals. *Mater. Today* **20**, 323 (2017).
3. I. A. Ovid'ko, R. Z. Valiev, Y. T. Zhu, Review on superior strength and enhanced ductility of metallic nanomaterials. *Prog.Mater. Sci.* **94**, 462 (2018).
4. J.-W. Yeh, S.-K. Chen, S.-J. Lin, J.-Y. Gan, T.-S. Chin, T.-T. Shun, C.-H. Tsau, S.-Y. Chang, Nanostructured High-Entropy Alloys with Multiple Principal Elements: Novel Alloy Design Concepts and Outcomes. *Adv. Eng. Mater.* **6**, 299 (2004).
5. B. Cantor, I. T. H. Chang, P. Knight, A. J. B. Vincent, Microstructural development in equiatomic multicomponent alloys. *Mater. Sci. Eng. A* **375-377**, 213 (2004).
6. E. P. George, D. Raabe, R. O. Ritchie, High-entropy alloys. *Nat. Rev.Mater.* **4**, 515 (2019).
7. Y. Zhang, T.T. Zuo, Z. Tang, M.C. Gao, K.A. Dahmen, P.K. Liaw, Z.P. Lu, Microstructures and properties of high-entropy alloys. *Prog.Mater. Sci.* **61**, 1 (2014).
8. Z. Li, S. Zhao, R.O. Ritchie, M.A. Meyers, Mechanical properties of high-entropy alloys with emphasis on face-centered cubic alloys. *Prog.Mater. Sci.* **102**, 296 (2019).
9. Z. Li, K.G. Pradeep, Y. Deng, D. Raabe, C.C. Tasan, Metastable high-entropy dual-phase alloys overcome the strength–ductility trade-off. *Nature* **534**, 227 (2016).
10. Z. Lei, X. Liu, Y. Wu, H. Wang, S. Jiang, S. Wang, X. Hui, Y. Wu, B. Gault, P. Kontis, D. Raabe, L. Gu, Q. Zhang, H. Chen, H. Wang, J. Liu, K. An, Q. Zeng, T.-G. Nieh, Z. Lu, Enhanced strength and ductility in a high-entropy alloy via ordered oxygen complexes. *Nature* **563**, 546 (2018).
11. Q. Ding, Y. Zhang, X. Chen, X. Fu, D. Chen, S. Chen, L. Gu, F. Wei, H. Bei, Y. Gao, M. Wen, J. Li, Z. Zhang, T. Zhu, R.O. Ritchie, Q. Yu, Tuning element distribution, structure and properties by composition in high-entropy alloys. *Nature* **574**, 223 (2019).
12. M.J. Yao, K.G. Pradeep, C.C. Tasan, D. Raabe, A novel, single phase, non-equiatomic FeMnNiCoCr high-entropy alloy with exceptional phase stability and tensile ductility. *Scr. Mater.* **72-73**, 5 (2014).
13. Z. Fu, W. Chen, H. Wen, D. Zhang, Z. Chen, B. Zheng, Y. Zhou, E.J. Lavernia, Microstructure and strengthening mechanisms in an FCC structured single-phase nanocrystalline Co25Ni25Fe25Al7.5Cu17.5 high-entropy alloy. *Acta Mater.* **107**, 59 (2016).
14. M.-H. Tsai, J.-W. Yeh, High-Entropy Alloys: A Critical Review. *Mater. Res. Lett.* **2**, 107 (2014).
15. A. Jain, S.P. Ong, G. Hautier, W. Chen, W.D. Richards, S. Dacek, S. Cholia, D. Gunter, D. Skinner, G. Ceder, K.A. Persson, Commentary: The Materials Project: A materials genome approach to accelerating materials innovation. *APL Mater.* **1**, 011002 (2013).
16. S. Curtarolo, G.L.W. Hart, M.B. Nardelli, N. Mingo, S. Sanvito, O. Levy, The high-throughput highway to computational materials design. *Nat. Mater.* **12**, 191 (2013).
17. D. Xue, P.V. Balachandran, J. Hogden, J. Theiler, D. Xue, T. Lookman, Accelerated search for materials with

targeted properties by adaptive design. *Nat. Comm.* **7**, 11241 (2016).

18. R. Ramprasad, R. Batra, G. Pilania, A. Mannodi-Kanakkithodi, C. Kim, Machine learning in materials informatics: recent applications and prospects. *npj Comput. Mater.* **3**, 54 (2017).
19. K.T. Butler, D.W. Davies, H. Cartwright, O. Isayev, A. Walsh, Machine learning for molecular and materials science. *Nature* **559**, 547 (2018).
20. D. Farkas, A. Caro, Model interatomic potentials and lattice strain in a high-entropy alloy. *J. Mater. Res.* **33**, 3218 (2018).
21. J. Li, Dislocation nucleation: Diffusive origins. *Nat. Mater.* **14**, 656 (2015).
22. L. Zhang, C. Lu, A. K. Tieu, Nonlinear elastic response of single crystal Cu under uniaxial loading by molecular dynamics study. *Mater. Lett.* **227**, 236 (2018).
23. B. Schuller, S. Steidl, A. Batliner, The INTERSPEECH 2009 emotion challenge. *INTERSPEECH* **2009**, 312 (2009).
24. G.-B. Huang, An Insight into Extreme Learning Machines: Random Neurons, Random Features and Kernels. *Cognitive Comput.* **6**, 376 (2014).
25. M. Dao, L. Lu, R. J. Asaro, J. T. M. De Hosson, E. Ma, Toward a quantitative understanding of mechanical behavior of nanocrystalline metals. *Acta Mater.* **55**, 4041 (2007).
26. Y. Mishin, M. Asta, J. Li, Atomistic modeling of interfaces and their impact on microstructure and properties. *Acta Mater.* **58**, 1117 (2010).
27. L. Zhang, C. Lu, K. Tieu, A review on atomistic simulation of grain boundary behaviors in face-centered cubic metals. *Comput.Mater. Sci.* **118**, 180 (2016).
28. J. R. Greer, J. T. M. De Hosson, Plasticity in small-sized metallic systems: Intrinsic versus extrinsic size effect. *Prog.Mater. Sci.* **56**, 654 (2011).
29. D. E. Spearot, M. A. Tschopp, K. I. Jacob, D. L. McDowell, Tensile strength of <1 0 0> and <1 1 0> tilt bicrystal copper interfaces. *Acta Mater.* **55**, 705 (2007).
30. M. A. Tschopp, D. L. McDowell, Influence of single crystal orientation on homogeneous dislocation nucleation under uniaxial loading. *J. Mech. Phys. Solids* **56**, 1806 (2008).
31. J. Schiøtz, K. W. Jacobsen, A Maximum in the Strength of Nanocrystalline Copper. *Science* **301**, 1357 (2003).
32. Q. Zhu, A. Samanta, B. Li, R. E. Rudd, T. Frolov, Predicting phase behavior of grain boundaries with evolutionary search and machine learning. *Nat.Comm.* **9**, 467 (2018).
33. C. Shen, C. Wang, X. Wei, Y. Li, S. van der Zwaag, W. Xu, Physical metallurgy-guided machine learning and artificial intelligent design of ultrahigh-strength stainless steel. *Acta Mater.* **179**, 201 (2019).
34. C.-T. Wu, H.-T. Chang, C.-Y. Wu, S.-W. Chen, S.-Y. Huang, M. Huang, Y.-T. Pan, P. Bradbury, J. Chou, H.-W. Yen, Machine learning recommends affordable new Ti alloy with bone-like modulus. *Mater. Today*. (2019).
35. S. Plimpton, Fast Parallel Algorithms for Short-Range Molecular Dynamics. *J. Comput. Phys.* **117**, 1 (1995).
36. A. Stukowski, Visualization and analysis of atomistic simulation data with OVITO–the Open Visualization Tool. *Modell. Simul.Mater.Sci.Eng.* **18**, 015012 (2010).
37. A. Stukowski, K. Albe, Dislocation detection algorithm for atomistic simulations, Modelling and Simulation in Materials Science and Engineering. *Modell. Simul.Mater.Sci.Eng.* **18**, 025016 (2010).

38. F. Wang, Y. Zhang, G. Chen, H. A. Davies, Tensile and Compressive Mechanical Behavior of a CoCrCuFeNiAl0.5 High Entropy Alloy. *Inter.J. Modern Phys. B* **23**, 1254 (2009).

## Acknowledgment

L.Z. and K.Q. would like to acknowledge the Postdoctoral Fellowship Program awarded by the Japan Society for the Promotion of Science (JSPS). We thank Satoru Fukuhara for assistance in constructing the database. This research is supported by the research grants from JSPS (P17711; P19081; 19H02415), the Start-up Funding for Young Faculty from Chongqing University, the Zhejiang Lab's International Talent Fund for Young Professionals (Project HANAMI), and the Australian Research Council Discovery Project (DP170103092).

## Author contribution

L.Z., K.Q., and Y.S. conceived the idea and initiated this project. L.Z. and C.L. build the models, carried out the molecular dynamic simulations, and constructed the database. L.Z. and K.Q. discussed and determined the machine learning scheme. K.Q. and B.W.S constructed and optimized the machine learning models. X.H., B.W.S, Y.S., and C.L. contributed to the fruitful discussions and supervision of the project. L.Z. and K.Q. prepared the manuscript, B.W.S conducted the proof reading. All authors discussed the results and commented on the manuscript.

## Additional information

**Supplementary Information** is submitted accompanies this paper.
**Competing interests:** The authors declare that they have no competing interests.

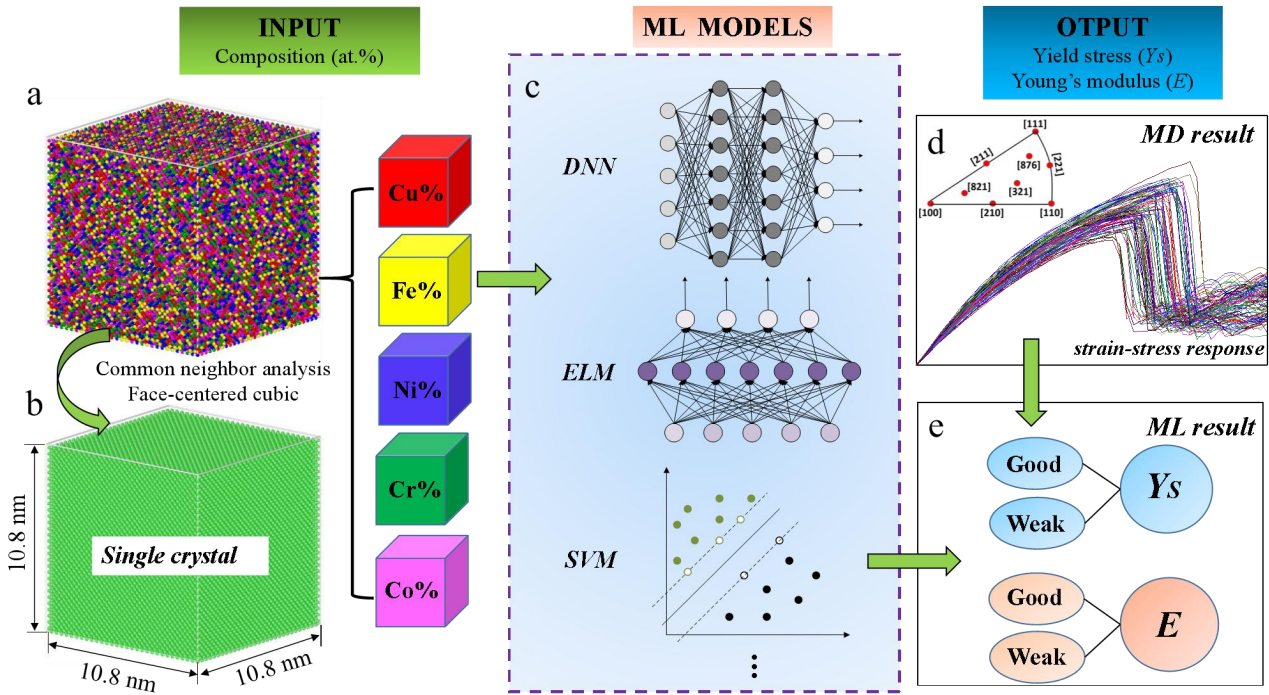

**Fig. 1. Schematic diagram of MD simulation and ML methods.** (**a**) Atomic configuration of a CuFeNiCrCo single-crystal sample, atoms are colored according to the element types. (**b**) Atoms are colored by the common neighbor analysis (CNA) method. The green atoms denote the face-centered cubic structure. (**c**) Schematic of the working principles of some ML models (DNN, ELM, and SVM). (**d**) Strain-stress response of single-crystal HEA samples with various element compositions along [110] orientation by MD simulation. The inverse pole figure indicates nine different crystallographic orientations tested in this study. (**e**) Prediction of the mechanical properties ($Ys$ and $E$) by ML method.

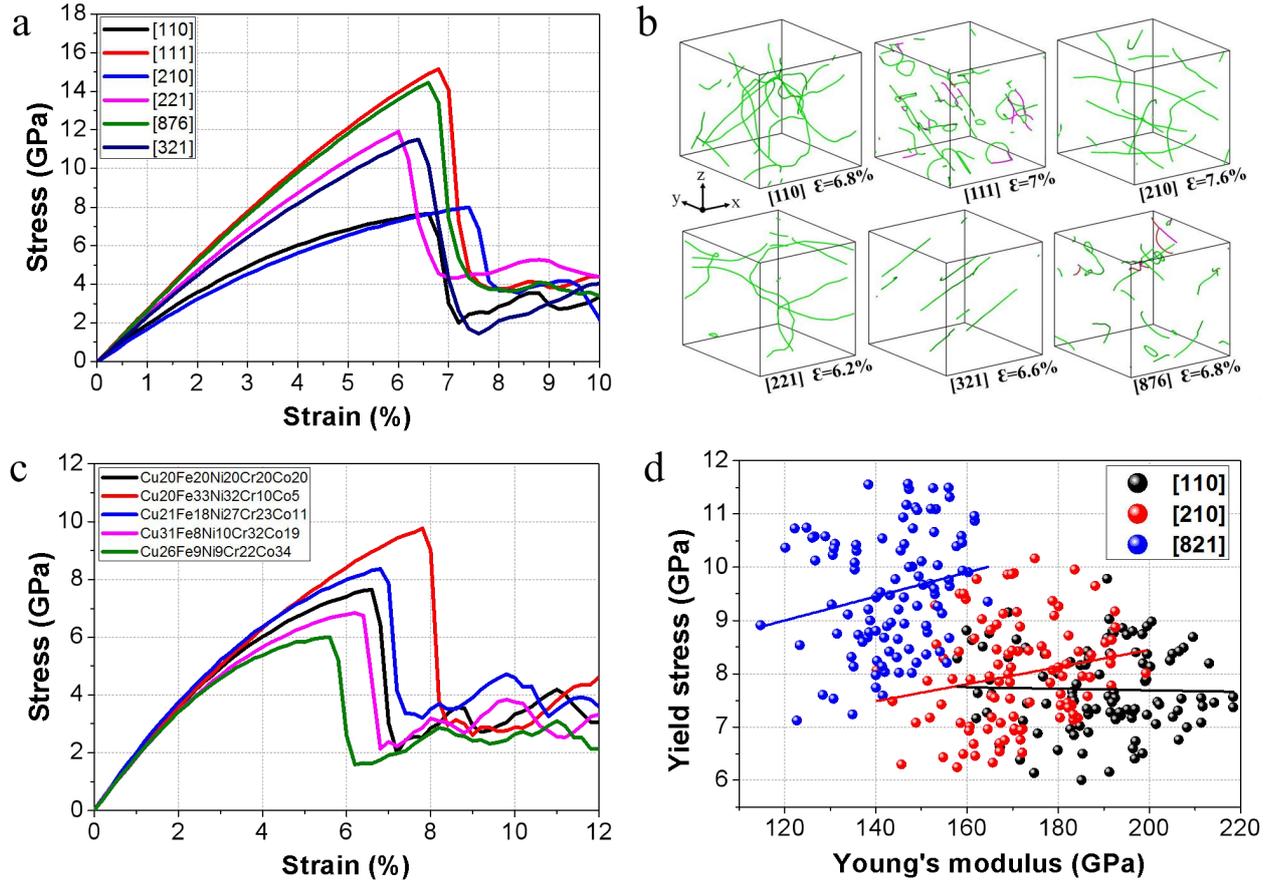

**Fig. 2. Mechanical responses of single-crystal HEA samples.** (**a**) Strain-stress response of the equiatomic single-crystal HEA samples with different crystallographic orientations. (**b**) Snapshots of dislocation nucleation in the single-crystal samples with different orientations near the yield point. The dislocations are extracted by the DXA algorithm, the different colors represent different types of dislocations. (**c**) Strain-stress response of the single-crystal HEA samples with different element compositions along [110] orientation. (**d**) Young's modulus as a function of the yield stress of single-crystal HEA samples with different element compositions along [110], [210], and [821] orientations.

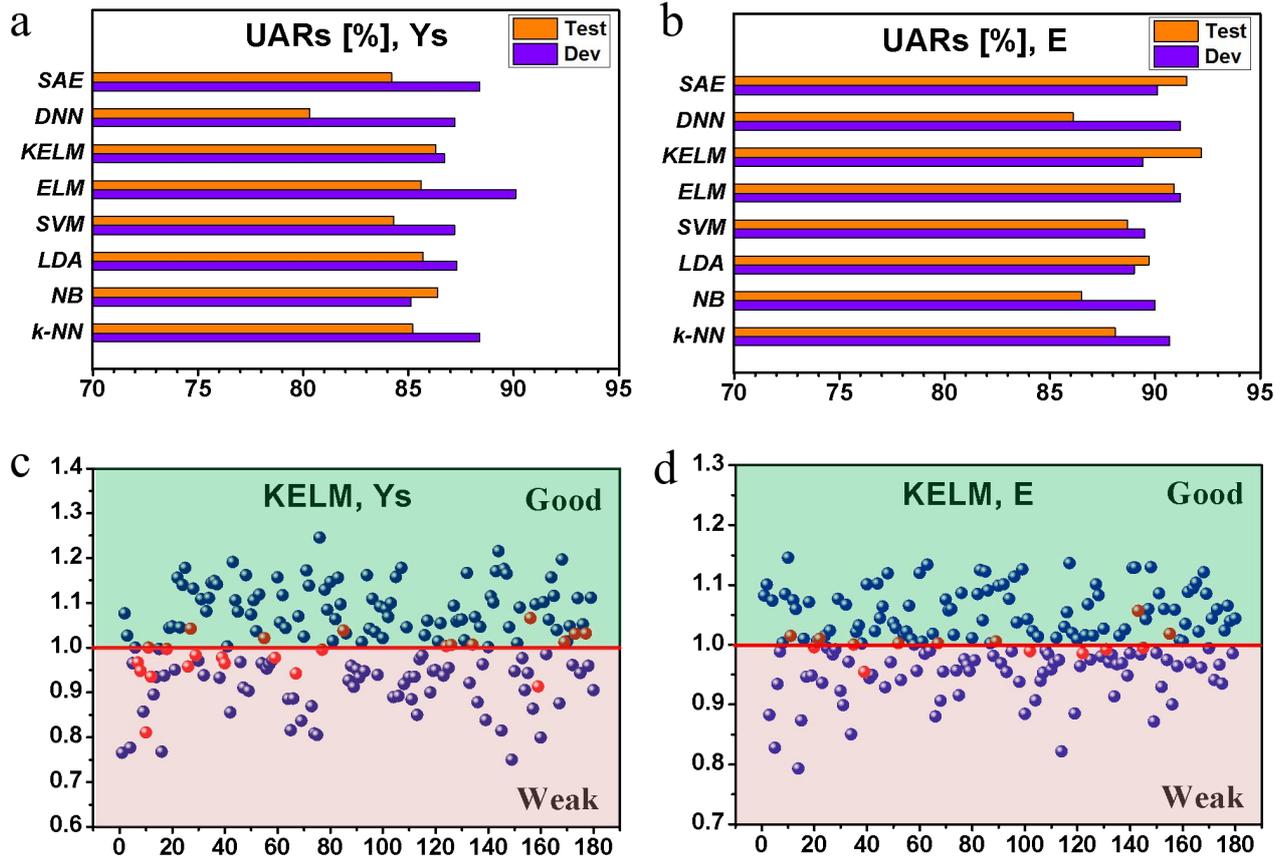

**Fig. 3. Results of ML models training and evaluation.** The results (UARs in [%]) of eight ML models on the task of (**a**) yield stress (*Ys)*, and (**b**) Young's modulus (*E*). The UARs shown for the dev set are the best UARs achieved by the optimal hyper-parameters tuned for the corresponding model. The UARs shown for the test set are the final performance achieved by the model trained by the train plus dev sets within the optimal hyper-parameters. Prediction result of the 180 non-equiatomic samples on the test set by KELM model for (**c**) yield stress (*Ys)*, and (**d**) Young's modulus (*E*). The red line is the benchmark line based on the value of the equiatomic sample, the results of non-equiatomic samples are normalized to the reference value, where the values above the line are classified to the 'Good' zone, and those below the line are classified to the 'Weak' zone. If the ML prediction matches the MD simulation, the point is colored blue, otherwise, the point is colored red.

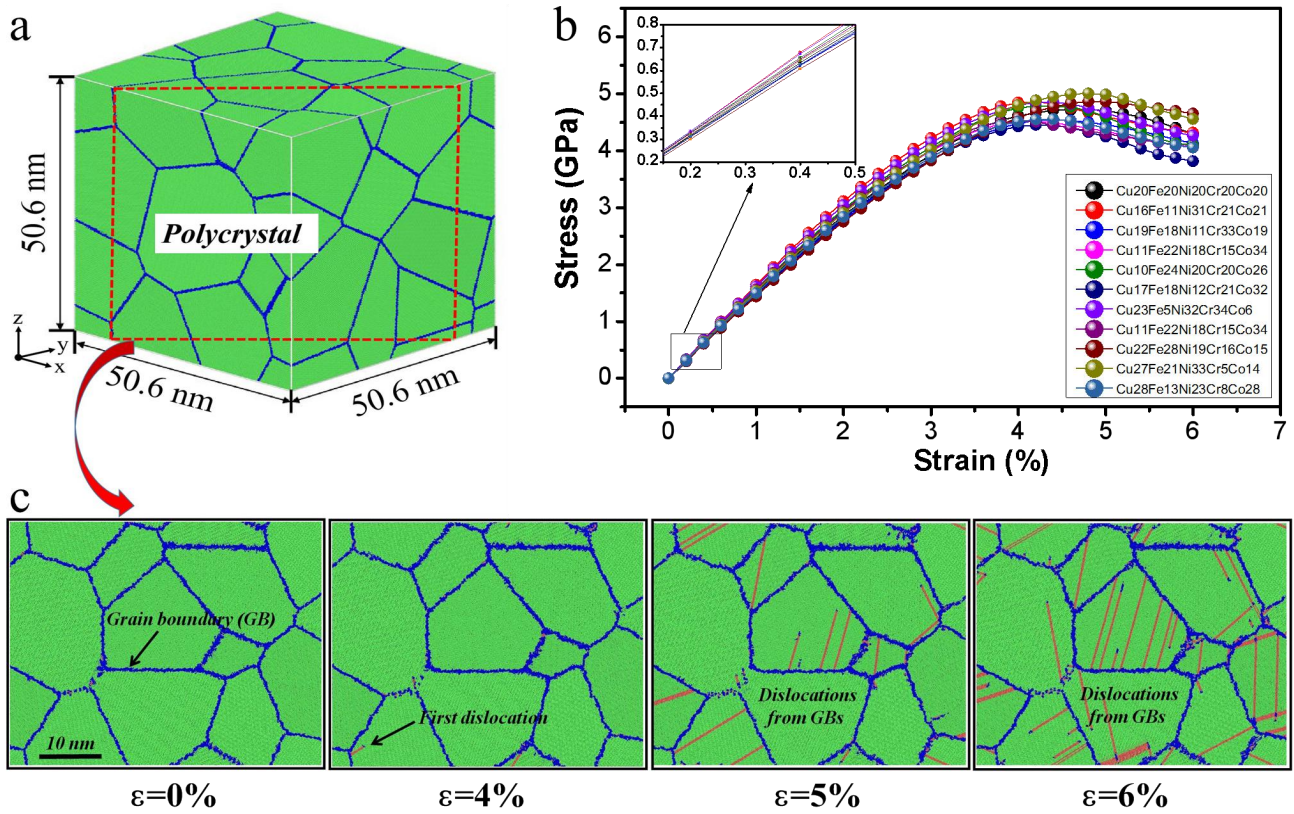

**Fig. 4. MD simulation of polycrystal HEA sample.** (**a**) Atomic configuration of the polycrystal sample. The sample contains 16 randomly orientated grains, and the mean grain size is 24.8 nm. (**b**) Stress-strain response of the polycrystal HEA samples with different element compositions. (**c**) Snapshots of the polycrystal sample $Cu_{28}Fe_{13}Ni_{23}Cr_8Co_{28}$ at different deformation stages under tension (cross-sectional view along [110] direction). Atoms are colored by the CNA method. The green atoms are located in the grain with fcc structure, the blue atoms are located at the grain boundary region with disordered structure, and the red atoms represent the stacking faults created by the slip of the nucleated dislocations from grain boundaries.

**Table 1. Prediction results of polycrystal samples by KELM model.** Comparison of MD simulation and ML prediction for yield stress and Young's modulus of ten polycrystal HEA samples with different element composition. '△' represents that the ML prediction is consistent with the MD simulation; otherwise, the ML result is marked as '×'.

| ID | Composition | Yield stress (GPa) | | Young's modulus (GPa) | |
|----|---|---|---|---|---|
| | | *MD simulation* | *ML result* | *MD simulation* | *ML result* |
| P0 | $Cu_{20}Fe_{20}Ni_{20}Cr_{20}Co_{20}$ | 4.72 | — | — | 150.99 | — | — |
| P1 | $Cu_{16}Fe_{11}Ni_{31}Cr_{21}Co_{21}$ | 4.86 | +2.97% | △ | 164.62 | +9.03% | △ |
| P2 | $Cu_{19}Fe_{18}Ni_{11}Cr_{33}Co_{19}$ | 4.55 | -3.60% | △ | 146.79 | -2.78% | × |
| P3 | $Cu_{11}Fe_{22}Ni_{18}Cr_{15}Co_{34}$ | 4.52 | -4.24% | △ | 156.26 | +3.49% | △ |
| P4 | $Cu_{10}Fe_{24}Ni_{20}Cr_{20}Co_{26}$ | 4.78 | 1.27% | × | 157.79 | +4.50% | △ |
| P5 | $Cu_{17}Fe_{18}Ni_{12}Cr_{21}Co_{32}$ | 4.45 | -5.72% | △ | 148.29 | -1.79% | × |
| P6 | $Cu_{23}Fe_{5}Ni_{32}Cr_{34}Co_{6}$ | 4.85 | +2.75% | △ | 161.30 | +6.83% | △ |
| P7 | $Cu_{11}Fe_{22}Ni_{18}Cr_{15}Co_{34}$ | 4.52 | -4.24% | △ | 156.28 | +3.50% | △ |
| P8 | $Cu_{22}Fe_{28}Ni_{19}Cr_{16}Co_{15}$ | 4.86 | +2.97% | △ | 144.09 | -4.57% | △ |
| P9 | $Cu_{27}Fe_{21}Ni_{33}Cr_{5}Co_{14}$ | 5.01 | +6.14% | △ | 153.48 | +1.65% | × |
| P10 | $Cu_{28}Fe_{13}Ni_{23}Cr_{8}Co_{28}$ | 4.55 | -3.60% | △ | 149.58 | -0.93% | × |

*Supplementary Materials for*

# Prediction on mechanical properties of non-equiatomic high-entropy alloy by atomistic simulation and machine learning

Liang Zhang[1,2,†,*], Kun Qian[3,†], Björn W. Schuller[4,5], Cheng Lu[6], Yasushi Shibuta[2,*], Xiaoxu Huang[1]

[1] College of Materials Science and Engineering, Chongqing University, Chongqing, 400044, China
[2] Department of Materials Engineering, The University of Tokyo, Tokyo, 113-8656, Japan
[3] Educational Physiology Laboratory, The University of Tokyo, Tokyo, 113-0033, Japan
[4] Department of Computing, Imperial College London, London, SW72AZ, UK
[5] Department of Computer Science, University of Augsburg, Augsburg, 86159, Germany
[6] Faculty of Engineering and Information Science, University of Wollongong, Wollongong, NSW 2522, Australia

[†] These authors contributed equally to this work.
[*] Corresponding author. Email: liangz@cqu.edu.cn (L.Z.); shibuta@material.t.u-tokyo.ac.jp (Y.S.)

## This PDF file includes:

Section S1. Calculation of the stacking fault energy by EAM potential.

Section S2. Phase stability of the HEA sample.

Section S3. Machine learning models.

Section S4. Evaluation Metrics.

Section S5. Hyper-parameters optimization.

Fig. S1. General stacking fault energy calculated by EAM potential.

Fig. S2. The HEA sample volume as a function of temperature.

Fig. S3. Tensile response of the single crystal HEA samples along different orientations.

Fig. S4. Young's modulus as a function of yield stress of the single crystal HEA samples.

Fig. S5. The hyper-parameters tuning process by different ML models.

Table S1. Basic characteristics of the five elements by CuFeNiCrCo EAM potential.

Table S2. Grid searching for optimizing the hyper-Parameters of the ML models.

Table S3. The optimal hyper-parameters selected for evaluating the ML models on the test set.

Table S4. The results of the ML models on the test set by complementary evaluation metrics.

## Section S1. Calculation of the stacking fault energy by EAM potential

For calculating general stacking fault (GSF) using MD, a simulation model was created with $[11\bar{2}]$, $[111]$ and $[1\bar{1}0]$ directions, and it was divided into two blocks in the normal direction. A free boundary condition was used in the normal direction ($[111]$ direction), while a periodic boundary condition was used in the lateral direction ($[11\bar{2}]$ and $[1\bar{1}0]$ direction). The GSF curve was determined by rigidly displacing the upper block on a $(111)$ plane along $[11\bar{2}]$ direction while fixing the lower block and calculating the energy change in the whole simulation model. Slip in $<112>$ direction is common, because stacking fault energy is lowest in this direction for the fcc system. Here, when displacing the upper block along $[11\bar{2}]$ direction, the lateral motion of atoms was constrained.

Along the path of displacement, the simulation model will have to first pass through an energy barrier that is referred as unstable stacking fault energy $\gamma_{usf}$. The displacement of the fcc lattice when $\gamma_{usf}$ reached equals to one-half of the partial Burgers vector $a_0/\sqrt{6}$ ($a_0$ is the equilibrium fcc lattice parameter). In the next stage, the simulation model became stable when the displacement is $a_0/\sqrt{6}$, although the model is not in its bulk equilibrium structure. The configuration now is known as the intrinsic or stable stacking fault $\gamma_{sf}$. Fig. S1 shows the GSF curve of Cu, Ni, and HEA samples by MD simulations using the CuFeNiCrCo EAM potential. The calculated $\gamma_{sf}$ values of Cu (45.6 mJ/m$^2$) and Ni (120.9 mJ/m$^2$) agree well with the reference values in Ref. [1]. The higher value of $\gamma_{usf}$ by MD simulation is mainly ascribed to the constrained boundary condition during the rigid displacement of the simulation sample. In general, the CuFeNiCrCo potential presents a large variation in stacking fault energies, both stable and unstable, which makes the interatomic potential useful to investigate the effects of a significant variation in the properties on mechanical behavior.

## Section S2. Phase stability of the HEA sample

The HEA samples (including both single-crystal sample and polycrystal sample) are constructed with a single fcc structural phase. HEAs with the fcc structure have been widely researched in previous experimental studies, and it has been found to be a generally stable phase of various HEAs [2]. Here, we further examine the phase stability of the CuFeNiCrCo sample by gradually raising the system temperature. A tested sample was constructed with a dimension of $10 \times 10 \times 10$ nm. It consists of the five elements arranged uniformly in a perfect fcc arrangement. The same interatomic potential was used here, as in the manuscript. Before the heating process, the HEA sample was firstly relaxed to an equilibrium configuration at 300 K in the canonical ensemble NVT (constant atom number, constant box volume, and constant temperature). Then, the temperature gradually increases from low temperature (300 K) to high temperature (3000 K) in the isothermal-isobaric ensemble (NPT). The HEA samples with five different element compositions were tested, including an equiatomic sample and four non-equiatomic samples. The total volume of the system was monitored during the heating process, as shown in Fig. S2. The inserted snapshots show the atomic configurations of the Cu$_{30}$Fe$_{18}$Ni$_9$Cr$_{14}$Co$_{29}$ sample at different temperature levels. Atoms are colored by the CNA method, where the green atoms are in fcc structure, and the blue atoms represent a disordered structure. It was found that the fcc phase of the HEA sample has not changed much below 1000 K. From 1500 K to 2000 K, the disordered atoms (blue atoms) increased considerably. The first-order phase transition (solid phase to liquid phase) was observed at about 2500 K, where the sample volume shows a sudden jump. Other simulated samples show a similar process although the melting temperature varies depending on the composition of the elements. The result shows that the fcc structure is a stable phase for CuFeNiCrCo alloy system.

## Section S3. Machine learning models

In this study, we investigate and compare eight ML models for the given learning targets, which include *shallow models* and the *deep models*. A brief description of each model is give as follows.

### *(1) Naïve Bayes*

Naïve Bayes (NB) [3] as classifier is based on a conditional probability model and the (usually naïve) assumption that features are independent of each other when given a class label. When performing as a classifier, the *maximum a posteriori* (MAP) algorithm is used, here. NB has been demonstrated to be a simple but efficient ML model in many tasks such as document classification, and spam filtering.

### *(2) Linear Discriminant Analysis*

Linear discriminant analysis (LDA) [4] originated from the *Fisher discriminant*, which bases on the assumption that the data generated by different classes underlies different Gaussian distributions (with mean and covariance parameters). In the training phase, LDA estimates the Gaussian distribution for each class. In the testing phase (to predict the class of a new sample), the trained LDA model will find the class which has the smallest misclassification cost.

### *(3) k-Nearest Neighbour*

A *k*-Nearest Neighbour (*k*-NN) [5] model, searches the *k* nearest neighbors to a given test instance from the train set in the feature space. Then, the prediction will be given to the test instance as the majority class variable of these *k* nearest neighbors. To find the nearest neighbors, *Euclidean distance* is usually used.

### *(4) Support Vector Machine*

A support vector machines (SVMs) [6] find a set of hyperplanes in a multi-dimensional space that instances belonging to different classes can be separated within. When training an SVM classier, a subset of data points with the widest possible gap between the class boundaries (based on the *support vectors*) from the train set will be selected as pivots to support the hyperplane that can maximize the separation between classes. This hyperplane is supposed to have the largest distance (*margin*) to the nearest train data points of any class. When testing a given instance, the test data (*feature vectors*) will be firstly mapped to the multi-dimensional space, then the prediction will be given based on which side of the gap they fall on to.

### *(5) Extreme Learning Machine*

An extreme learning machine (ELM) [7] is a kind of a single hidden layer feed-forward artificial neural network (FNN), which has no need for tuning the parameters of the network. For an ELM, the hidden neurons are randomly initialized, by which the output weights can be analytically determined. Unlike the backpropagation method used in training 'classic' FNN, ELM only use the Moore-Penrose generalized inverse of the hidden layer matrix to estimate the output weights. Therefore, an ELM can be very fast in both, the training and testing phases.

*(6) Kernel-based Extreme Learning Machine*

If the feature mapping is unknown, then the kernel trick used in SVM can also be applied to ELM. This model is known as kernel-based ELM (KELM) [8]. In this study, KELM could outperform SVM when kernels are used due to the fact that an SVM may tend to reach sub-optimal solutions.

*(7) Deep Neural Network*

The principle of deep learning (DL) [9] is to extract higher representations from the data via the help of a series of nonlinear transformation of the inputs. There are several types of DNN architecture, e.g., the convolutional neural network (CNN), or the recurrent neural network (RNN). In this paper, DNN refers to a FNN with multiple hidden layers (four in this study). Each layer is fully connected with all neurons of the subsequent layer, whereas there are no connections between nodes within the same layer, or across the multiple layers. The procedure of training a DNN is to update the parameters of its layers (weights and biases) iteratively for minimizing a pre-defined loss function, which measures the difference between the target output vectors and the actual output vectors of the network.

*(8) Stacked Autoencoders*

Stacked autoencoders (SAEs) [10] are a kind of efficient DL models that can learn higher representations from the inputs via an unsupervised learning paradigm in the initialization of the network. An autoencoder (AE) is usually an FNN model (often with one hidden layer) having an equal number of neurons between the inputs and the outputs. The training process of an AE (composed of an *encoder* and a *decoder*) is to reconstruct the inputs in the output layer. The optimization process of training an AE is similar to training a DNN as mentioned above. In this study, we also implement the *Scaled Conjugate Gradient* (SCG) method during the training process for AE. When building the SAE model, AEs are stacked together one layer by one layer, by which the learnt features from lower a layer can be regarded as inputs of the subsequent higher layer. Once all layers are pre-trained, a fine-tuning process can be executed to finish by a supervised learning process.

## Section S4. Evaluation Metrics

Due to the imbalanced characteristic of our proposed database, we use the unweighted average recall (UAR) as our main metric to evaluate the model's performance. Firstly, let us define the *recall* (aka. class-wise accuracy) of the $i$-th class as:

$$Recall_i = \frac{\overline{N_i}}{N_i},$$

where $\overline{N_i}$ and $N_i$ are the correctly recognized instance number and the total instance number of the $i$-th class,

respectively. The weighted average recall (WAR) is defined as:

$$\text{WAR} = \sum_{i=1}^{N_c} \lambda_i \, Recall_i$$

$$\lambda_i = \frac{N_i}{N},$$

where $N_c$ and $N$ are the number of classes in the task and the total number of instances, respectively. We know that, if the data distribution is extremely imbalanced, an ML model will usually be trained much stronger in recognizing the classes which occupy a larger percentage than the other classes in the total dataset. Therefore, using WAR (or accuracy) to evaluate the final performance for an imbalanced dataset could be overoptimistic. In contrast, the UAR is defined as:

$$\text{UAR} = \frac{\sum_{i=1}^{N_c} Recall_i}{N_c}.$$

We can see that, for a balanced dataset (i.e., $\lambda_i$ is a constant), WAR is equal to UAR. Considering our task is a binary classification problem, we also provide a series of complementary evaluation metrics. They are defined as:

$$\text{sensitivity} = \frac{TP}{TP+FN}$$

$$\text{specificity} = \frac{TN}{TN+FP}$$

$$\text{precision} = \frac{TP}{TP+FP}$$

$$\text{F1 score} = 2 \cdot \frac{\text{precision} \cdot \text{sensitivity}}{\text{precision} + \text{sensitivity}} = \frac{2TP}{2TP+FP+FN},$$

where TP, TN, FP, and FN are the number of *true positive* ("Good" correctly identified as "Good"), *true negative* ("Weak" correctly identified as "Weak"), *false positive* ("Weak" incorrectly identified as "Good"), and *false negative* ("Good" incorrectly identified as "Weak"), respectively.

## Section S5. Hyper-parameters optimization

The hyper-parameters of all ML models are tuned and optimized on the dev set based on the performance (UAR). We use a grid-search strategy to decide on the optimal hyper-parameter for a specific ML model. All the ML models (except SVM) are implemented by the MATLAB (R2019a) *Statistics and Machine Learning Toolbox*, MATLAB (R2019a) *Deep Learning Toolbox*, and MATLAB (K)ELM Codes. The SVM model is implemented by the LIBSVM Toolkit. We use the default hyper-parameters designed in these tools except if claimed as follows.

Table S2 shows the grid search strategy we use for optimizing the hyper-parameters of the ML models. From Fig. S5 we can see that, optimization of the hyper-parameters is necessary for building an efficient ML model. All the hyper-parameters of the models are tuned and optimized based on their performance (UARs) on the dev set.

Then, the optimal hyper-parameter will be applied to train a new model based on the combination of the train and dev sets. Finally, the test set (unseen) will be evaluated by this new model.

Table S3 shows the optimal hyper-parameters selected for evaluating the final ML models on the test set. When tuning the hyper-parameters by the grid search strategy, the earliest value (calculating from the start point of the grid) will be selected if there is more than one peak value (i.e., highest UAR) achieved in the whole grid.

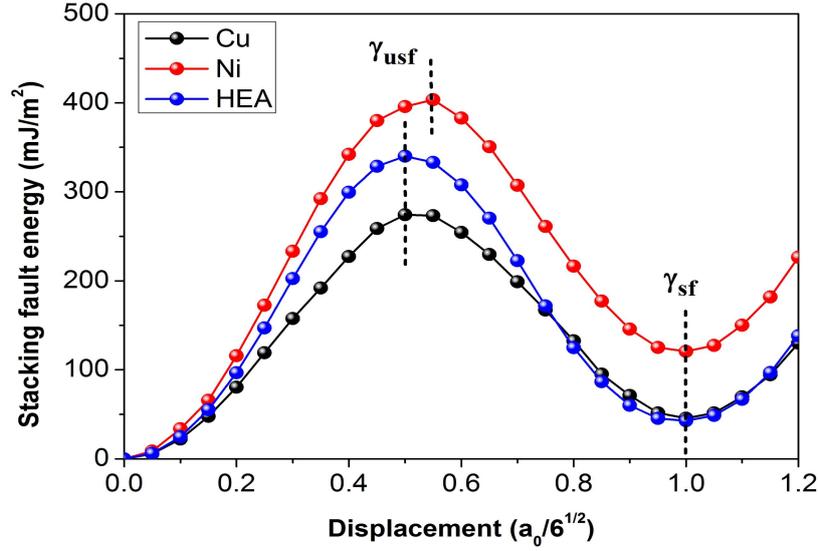

**Fig. S1.** The general stacking fault energy curves of Cu, Ni, and HEA using the CuFeNiCrCo EAM potential. $\gamma_{usf}$ is the unstable stacking fault energy, and $\gamma_{sf}$ is the stacking fault energy.

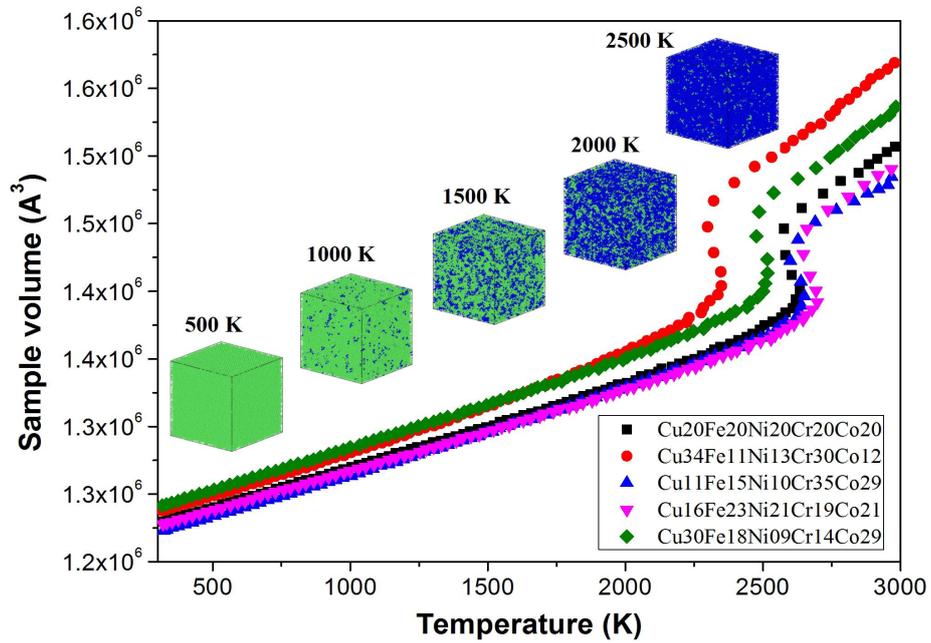

**Fig. S2.** The HEA sample volume as a function of temperature from 300 K to 3000 K. The inserted snapshots show the atomic configurations of the $Cu_{30}Fe_{18}Ni_9Cr_{14}Co_{29}$ sample at different temperature levels.

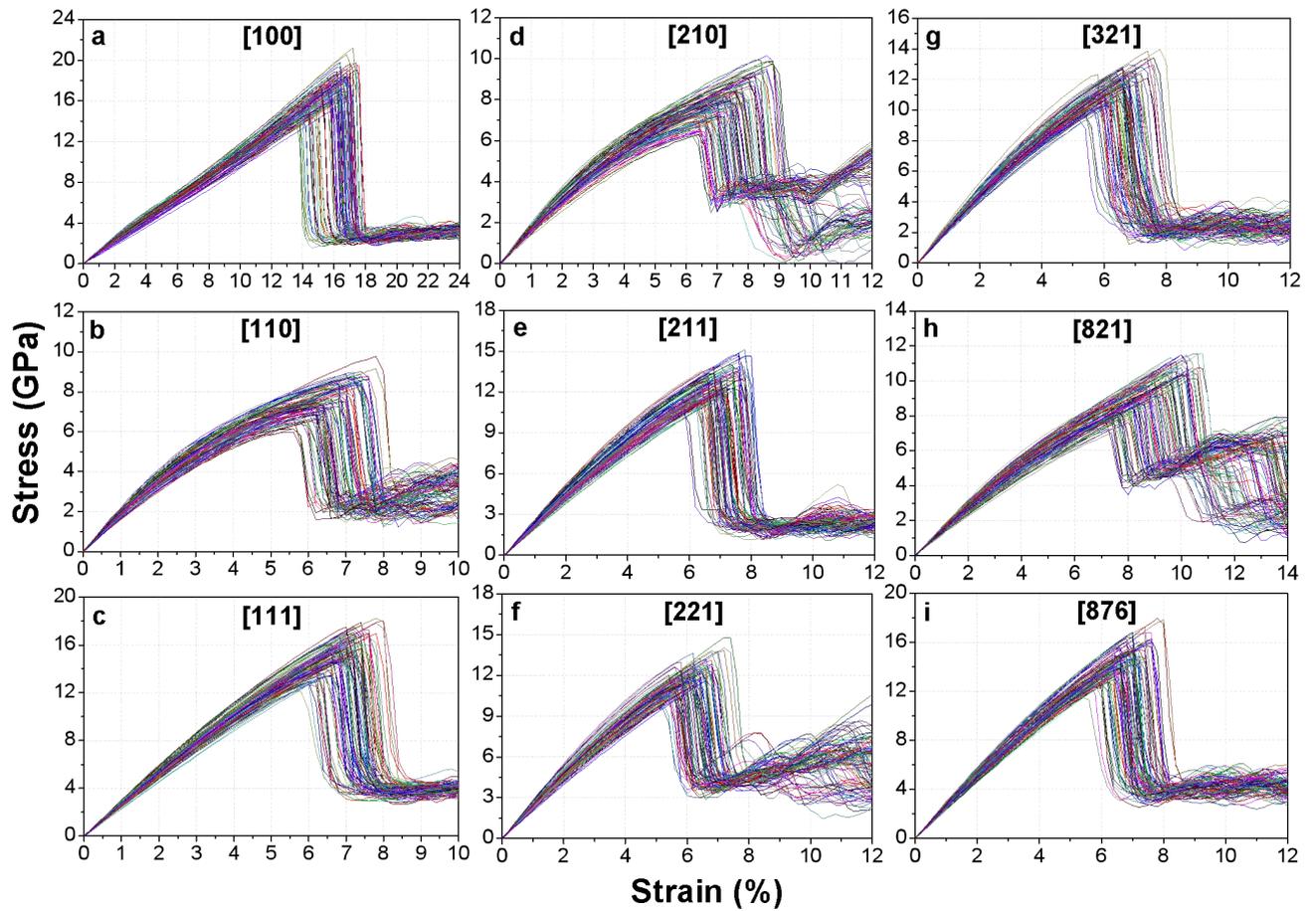

**Fig. S3.** Stress-strain response of the single crystal HEA samples with various element compositions along different orientations. Each branch figure contains the result of one equiatomic sample and 100 non-equiatomic samples.

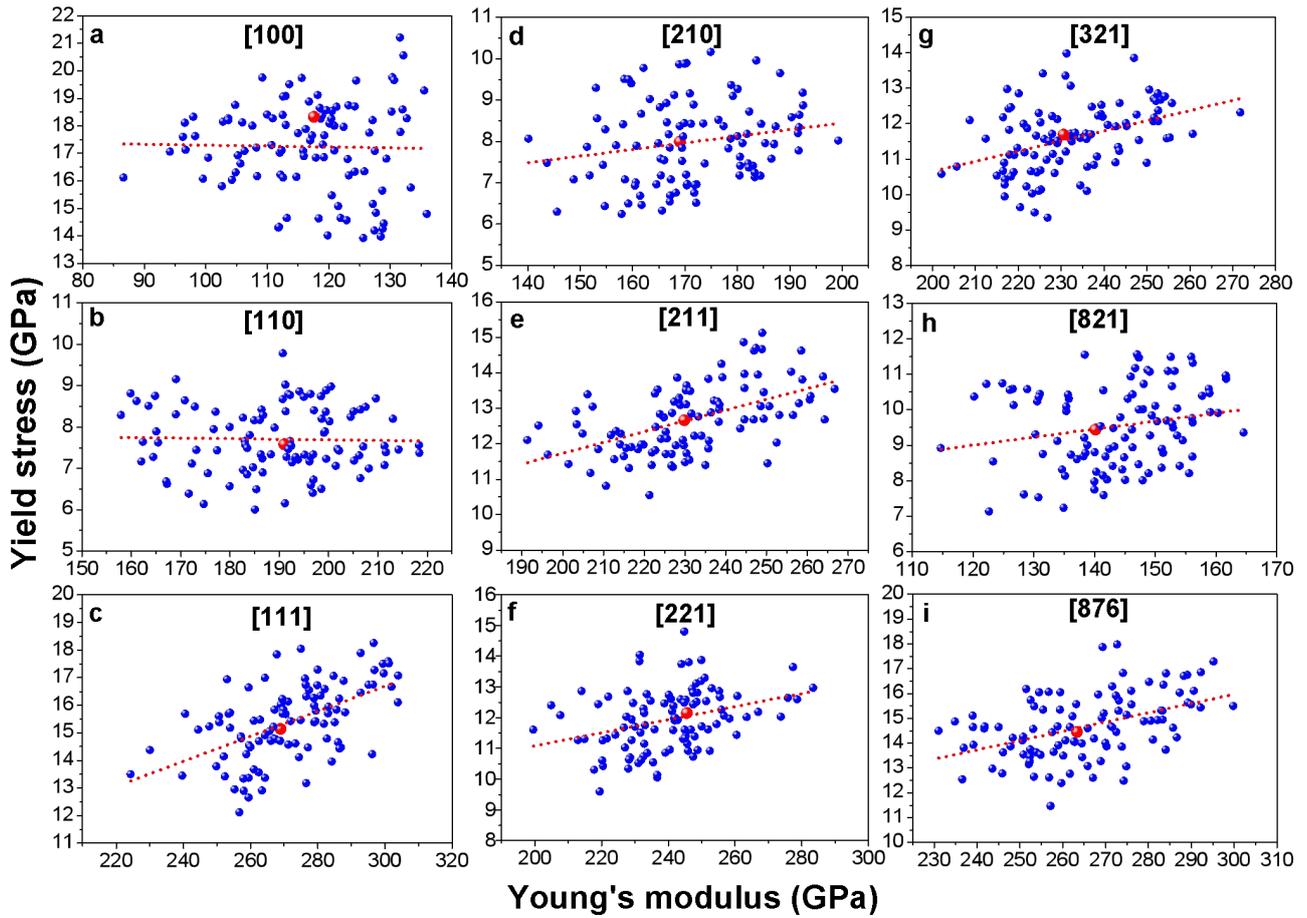

**Fig. S4.** Young's modulus as a function of yield stress of single crystal HEA samples with various element compositions along different orientations. The red point denotes the result of the equiatomic sample.

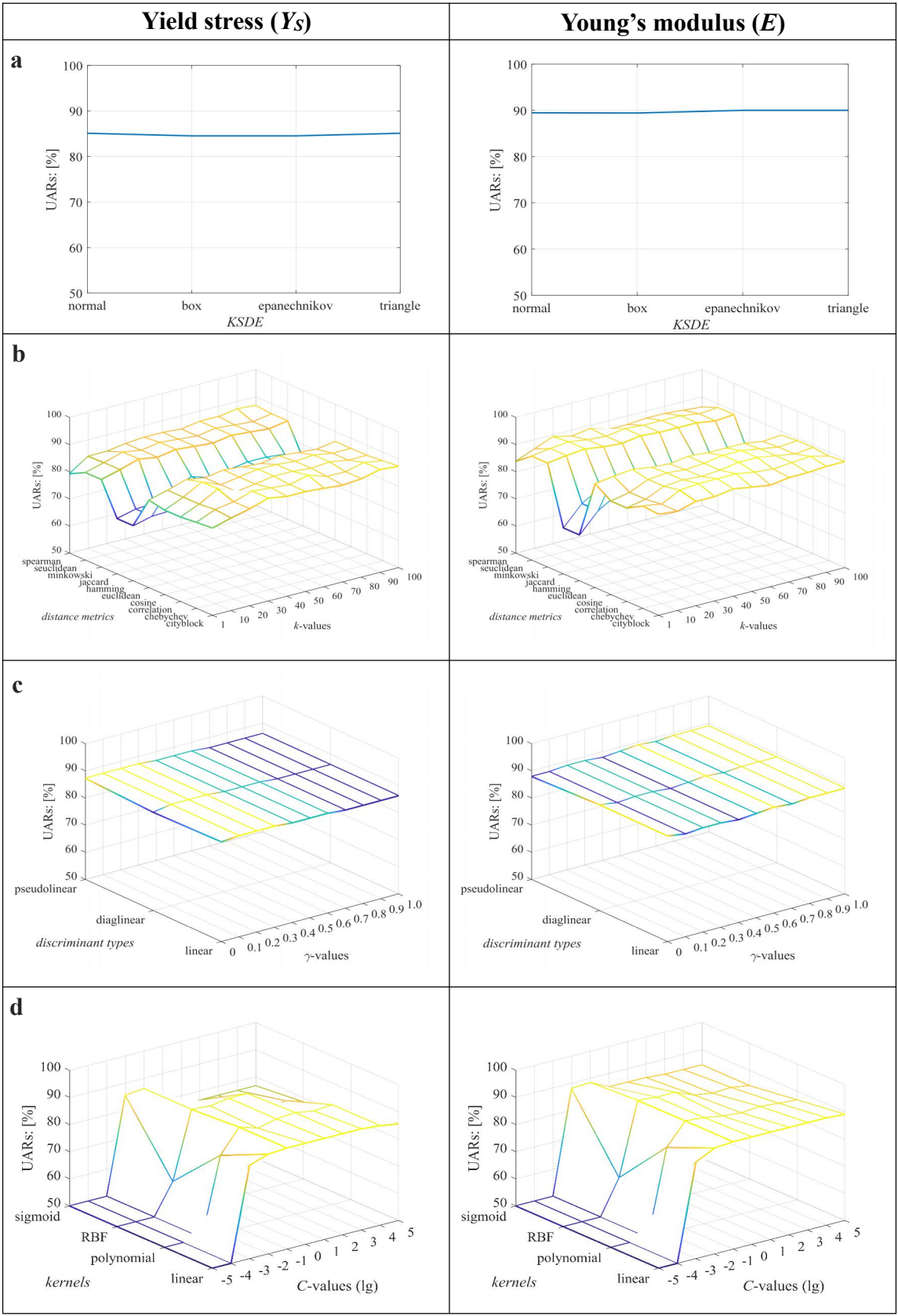

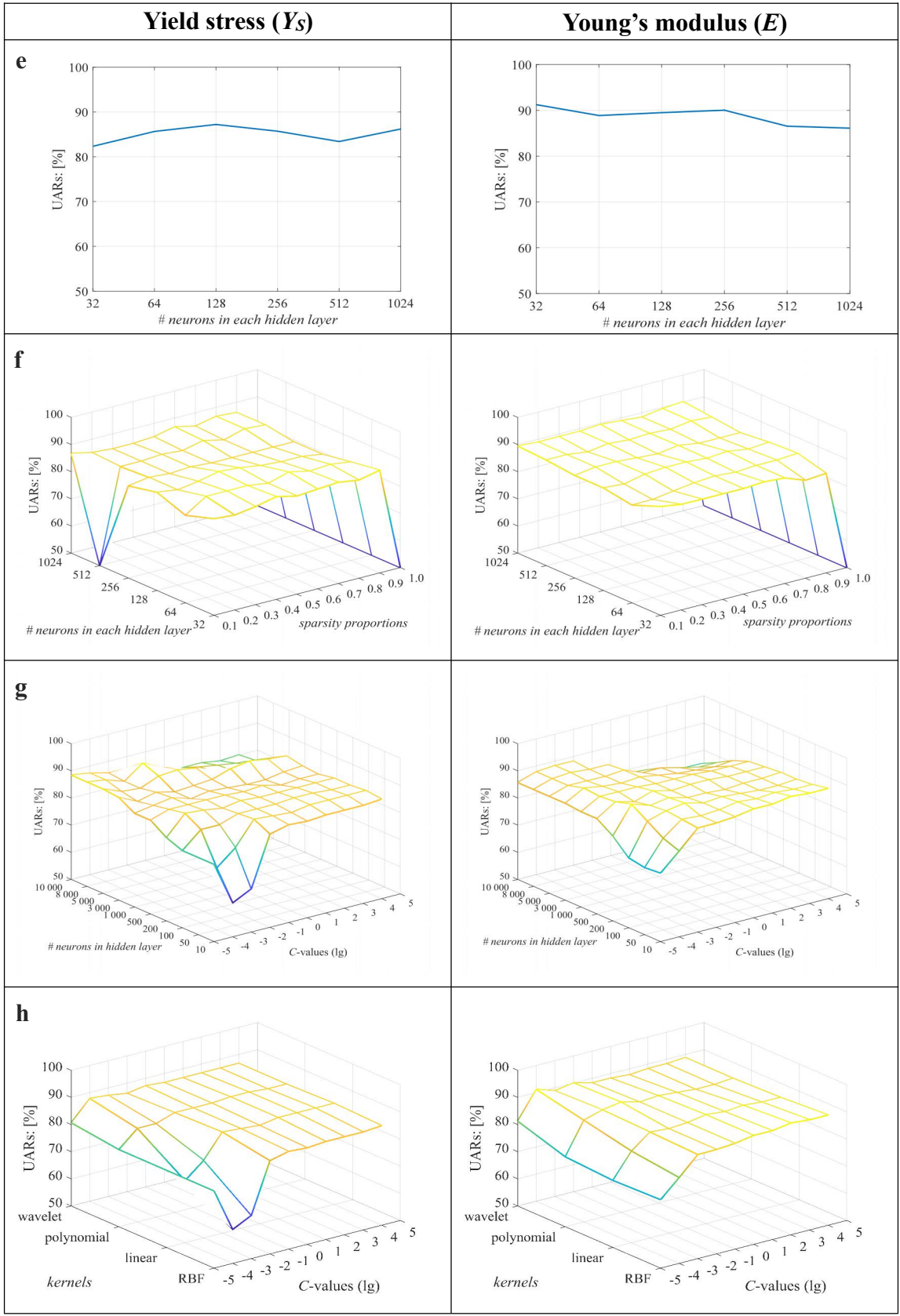

**Fig. S5.** The hyper-parameters tuning process for yield stress and the Young's modulus by different ML models. (a) NB model, (b) *k*-NN model, (c) LDA model, (d) SVM model, (e) DNN model, (f) SAE model, (g) ELM model, and (h) KELM model.

**Table S1** Basic characteristics of the five constituent elements (CuFeNiCrCo) of the HEA sample, including lattice constant ($a_0$), cohesive Energy ($E_{coh}$), elastic constants ($C_{11}$, $C_{12}$, $C_{44}$), stable stacking fault energy ($\gamma_{sf}$), and unstable stacking fault energy ($\gamma_{usf}$).

| Elements | $a_0$ (Å) | $E_{coh}$ (eV) | Elastic Constants (GPa) | | | $\gamma_{sf}$ (mJ/m²) | $\gamma_{usf}$ (mJ/m²) |
|---|---|---|---|---|---|---|---|
| | | | $C_{11}$ | $C_{12}$ | $C_{44}$ | | |
| Cu | 3.62 | 3.54 | 106 | 76 | 48 | 46* / 45.6** | 190 / 274 |
| Fe | 3.56 | 4.40 | 119 | 100 | 48 | 43 | 328 |
| Ni | 3.52 | 4.45 | 154 | 92 | 78 | 129 / 120.9 | 363 / 403 |
| Cr | 3.53 | 4.20 | 124 | 88 | 70 | 41 | 261 |
| Co | 3.55 | 4.41 | 165 | 120 | 89 | 41 | 129 |

\* Value from Ref. [1]

\*\* Value by MD simulation.

**Table S2.** Grid search for optimizing the hyper-Parameters of the ML models.

| ML Models | Hyper-Parameters |
|---|---|
| NB | *kernel smoothing density estimate (KSDE)*: {"normal", "box", "epanechnikov", "triangle"}. |
| *k*-NN | *k*-values: {1, 10, 20, 30, 40, 50, 60, 70, 80, 90, 100}; *distance metric*: {"cityblock", "chebychev", "correlation", "cosine", "euclidean", "hamming", "jaccard", "minkowski", "seuclidean", "spearman"}. |
| LDA | *discriminant types*: {"linear", "diaglinear", "pseudolinear"}; $\gamma$-values: {0.0, 0.1, 0.2, 0.3, 0.4, 0.5, 0.6, 0.7, 0.8, 0.9, 1.0}. |
| SVM | *kernel types*: {"linear", "polynomial", "RBF", "sigmoid"}; *C-values*: {$10^{-5}$, $10^{-4}$, $10^{-3}$, $10^{-2}$, $10^{-1}$, $10^{0}$, $10^{1}$, $10^{2}$, $10^{3}$, $10^{4}$, $10^{5}$}. |
| DNN | *4 hidden layers*: {[32-32-32-32], [64-64-64-64], [128-128-128-128], [256-256-256-256], [512-512-512-512], [1024-1024-1024-1024]} |
| SAE | *2 hidden layers* (# neurons): {[32-32], [64-64], [128-128], [256-256], [512-512], [1024-1024]}; *sparsity proportions*: {0.1, 0.2, 0.3, 0.4, 0.5, 0.6, 0.7, 0.8, 0.9, 1.0}; *L2 weight regularisation*: 0.0001; *sparsity regularisation*: 2. |
| ELM | *activation function*: "sigmoid"; *hidden layer sizes* (# neurons): {10, 50, 100, 200, 500, 1 000, 3 000, 5 000, 8 000, 10 000}; *C-values*: same as in SVM. |
| KELM | *kernel types*: {"linear", "polynomial", "RBF", "wavelet"}; *C-values*: same as in SVM; *kernel parameters*: [1, 10, 100]. |

**Table S3.** The optimal hyper-parameters selected for evaluating the ML models on the test set.

| ML Models | | *Ys* | *E* |
|---|---|---|---|
| **NB** | *KSDE* | "normal" | "epanechnikov" |
| **k-NN** | *k-value* | 60 | 50 |
| | *distance metric* | "chebychev" | "chebychev" |
| **LDA** | *discriminant type* | "pseudolinear" | "linear" |
| | *γ-value* | 0 | 0.6 |
| **SVM** | *kernel type* | "polynomial" | "linear" |
| | *C-value* | 0.1 | 0.1 |
| **DNN** | *# neurons in each hidden layer* | 128 | 32 |
| **SAE** | *sparsity proportion* | 0.6 | 0.9 |
| | *# neurons in each hidden layer* | 64 | 1024 |
| **ELM** | *# neurons in hidden layer* | 200 | 50 |
| | *C-value* | 1 000 | 0.01 |
| **KELM** | *kernel type* | "linear" | "RBF" |
| | *C-value* | 0.00001 | 0.1 |

**Table S4.** The results of the ML models on the test set by complementary evaluation metrics (in [%]).

| ML model | Task | WAR | Sens. | Spec. | Prec. | F1-score |
|---|---|---|---|---|---|---|
| NB | $Y_s$ | 86.7 | 88.9 | 84.0 | 87.1 | 88.0 |
| NB | $E$ | 86.7 | 88.0 | 85.0 | 88.0 | 88.0 |
| $k$-NN | $Y_s$ | 85.6 | 88.9 | 81.5 | 85.4 | 87.1 |
| $k$-NN | $E$ | 88.3 | 90.0 | 86.3 | 89.1 | 89.6 |
| LDA | $Y_s$ | 86.1 | 89.9 | 81.5 | 85.6 | 87.7 |
| LDA | $E$ | 90.0 | 92.0 | 87.5 | 90.2 | 91.1 |
| SVM | $Y_s$ | 84.4 | 85.9 | 82.7 | 85.9 | 85.9 |
| SVM | $E$ | 88.9 | 90.0 | 87.5 | 90.0 | 90.0 |
| DNN | $Y_s$ | 80.6 | 82.8 | 77.8 | 82.0 | 82.4 |
| DNN | $E$ | 86.1 | 86.0 | 86.3 | 88.7 | 87.3 |
| SAE | $Y_s$ | 84.4 | 86.9 | 81.5 | 85.1 | 86.0 |
| SAE | $E$ | 91.7 | 93.0 | 90.0 | 92.1 | 92.5 |
| ELM | $Y_s$ | 86.1 | 90.9 | 80.2 | 84.9 | 87.8 |
| ELM | $E$ | 90.6 | 88.0 | 93.8 | 94.6 | 91.2 |
| KELM | $Y_s$ | 86.7 | 89.9 | 82.7 | 86.4 | 88.1 |
| KELM | $E$ | 92.2 | 92.0 | 92.5 | 93.9 | 92.9 |

**References**

[1] D. Farkas, A. Caro. Model interatomic potentials and lattice strain in a high-entropy alloy, Journal of Materials Research 33 (2018) 3218-3225.

[2] Z. Li, S. Zhao, R.O. Ritchie, M.A. Meyers. Mechanical properties of high-entropy alloys with emphasis on face-centered cubic alloys, Progress in Materials Science 102 (2019) 296-345.

[3] M.N. Murty, V.S. Devi, (Eds.). Pattern Recognition: An Algorithmic Approach, Netherlands: Springer Science & Business Media, Dordrecht, 2011.

[4] R.A. Fisher. The use of multiple measurements in taxonomic pronlems Annals of Eugenics 7 (1936) 179-188.

[5] T. Cover, P. Hart. Nearest neighbor pattern classification, IEEE Transactions on Information Theory 13 (1967) 21-27.

[6] C. Cortes, V. Vapnik. Support-vector networks, Machine Learning 20 (1995) 273-297.

[7] G.-B. Huang, Q.-Y. Zhu, C.-K. Siew. Extreme learning machine: Theory and applications, Neurocomputing 70 (2006) 489-501.

[8] G. Huang, H. Zhou, X. Ding, R. Zhang. Extreme Learning Machine for Regression and Multiclass Classification, IEEE Transactions on Systems, Man, and Cybernetics, Part B (Cybernetics) 42 (2012) 513-529.

[9] Y. LeCun, Y. Bengio, G. Hinton. Deep learning, Nature 521 (2015) 436-444.

[10] P. Vincent, H. Larochelle, I. Lajoie, Y. Bengio, P.-A. Manzagol. Stacked Denoising Autoencoders: Learning Useful Representations in a Deep Network with a Local Denoising Criterion, J. Mach. Learn. Res. 11 (2010) 3371–3408.